# Direct determination of one-dimensional interphase structures using normalized crystal truncation rod analysis

Tomoya Kawaguchi[1], Yihua Liu[1], Anthony Reiter[2], Christian Cammarota[2], Michael S. Pierce[2], and Hoydoo You[1]

[1]*Materials Science Division, Argonne National Laboratory, Argonne, Illinois, 60439*

[2]*Rochester Institute of Technology, School of Physics and Astronomy, Rochester NY, 14623*

A one-dimensional non-iterative direct method was employed for normalized crystal truncation rod. The non-iterative approach, utilizing Kramers-Kronig relation, avoids the ambiguities due to the improper initial model or the incomplete convergence in the conventional iterative methods. The validity and limitation of the present method are demonstrated through both numerical simulations and experiments with Pt (111) in 0.1 M CsF aqueous solution. The present method is compared to conventional iterative phase-retrieval methods.

# 1. Introduction

Understanding structures of interphases, extended phases of an interface, are of great importance in various disciplines. The field of electrochemistry is particularly impacted because properties of energy systems such as batteries and fuel cells are dominated by interphase structures both in kinetics and thermodynamics (Winter & Brodd, 2004). Crystal truncation rod (CTR) analysis is a powerful tool for investigating interphase structures with atomic resolution and has played an important role in the electrochemical interface studies (Nagy & You, 2002). Because the phases of CTR amplitudes are lost in measurements, modeling is conventionally employed to determine the real structure. The reliability of the obtained result is, however, often limited because of inevitable dependence on the initial models employed. Thus, the development of a model-independent, direct analysis method is desirable for the CTR analysis by solving the phase problem.

There have been numerous studies for solving the phase problem. Applying and extending optical techniques (Fienup, 1982), Miao *et al.* demonstrated that the lost phase can be iteratively retrieved from the scattering intensity by oversampling the reciprocal space with a *support* in the real space where electron density exists (Miao et al., 1999). Similarly, iterative phase-retrieval methods for CTR and surface X-ray diffraction have been also developed (Saldin & Shneerson, 2008) aiming for retrieval of the three-dimensional electron density distribution. The examples include COBRA (Yacoby et al., 2000), PARADIGM (Fung et al., 2007) and DCAF (Björck et al., 2008), in which the models converge iteratively due to the imposition of constraints such as positive electron density, finite thickness of an interphase structure, and so on. These model-independent iterative methods are advantageous to the modeling. However, there still exists a possible ambiguity whether the calculation has attained the true minimum (Werner et al., 2010; Pauli et al., 2012). Furthermore, the constraints imposed in the iterative methods may not be available for a diffuse electron density in buried interfaces where the methods have to handle a negative electron density or a semi-infinite electron density. Resonant X-ray scattering can be used to determine the phase (Pauli et al., 2012). However, the resonant scattering measurement is not possible without resonant atoms in the structure. Thus, a direct analysis method without the need of the constraints would be useful for the CTR analysis.

In the present study, a non-iterative one-dimensional (1D) direct inversion method is developed by using the causality relationship, known as the Hilbert transform in mathematics and as the Kramers-Kronig relation (KKR) in physics. This relation enables to calculate an imaginary part of a structure factor from the corresponding real part and vice versa. Therefore, it is possible to solve directly the phase problem in the 1D CTR.

This paper is organized as follows. First, the theory of the present method is described. Second, numerical simulations are performed to demonstrate the validity of the present method.

Third, the method was applied for the experimental data obtained from the system of Pt(111) in CsF electrolyte. Finally, we discuss the 'causality' condition of KRR of our method and other potential applications, followed by a conclusion.

## 2. Method

### 2.1 Theory

The CTR intensity is proportional to $|F + f|^2$ where $F$ and $f$ are the substrate and interphase structure factors, respectively. Then, the normalized CTR, $\bar{I}$, which is free from geometric corrections, is expanded to

$$\frac{|F|}{2}(\bar{I} - 1) = \frac{F'}{|F|}f' + \frac{F''}{|F|}f'' + \frac{f'^2 + f''^2}{2|F|}, \qquad (1)$$

where $\bar{I} \equiv |F + f|^2/|F|^2$, $F \equiv F' + iF''$, and $f \equiv f' + if''$. In order to solve directly this equation for $f'$ and $f''$ from known $\bar{I}$ and $F$, we will employ the Hilbert transformation,

$$f' = -H[f''], \qquad (2)$$

where the Hilbert transform, $H[f'']$, is defined using Cauchy principal value, $P$, as

$$H[u(k)] \equiv \frac{1}{\pi} P \int_{-\infty}^{\infty} \frac{u(k')}{k - k'} \mathrm{d}k'. \qquad (3)$$

Strictly speaking, Eq. (2) is valid when the interphase structure exists only in the positive coordinate. Since there are two equations and two unknowns, $f'$ and $f''$ can be solved numerically. The electron density distribution of the interphase $\rho = \mathcal{F}^{-1}[f' + if'']$ where $\mathcal{F}^{-1}$ denotes the inverse Fourier transform.

The mathematical definition of KKR is derived from the causality; an effect must happen *after* a cause in the time domain. The time domain causality is widely used in resonance scattering and spectroscopy. The energy spectrum of the real part of a resonant scattering term is calculated from its imaginary part and vice versa using KKR. It also enables us to retrieve a phase of a complex scattering factor from a modulus in the frequency domain using the logarithmic dispersion relation (Roessler, 1965; Kawaguchi et al., 2014, 2017). The present study applies the causality of KKR to the spatial domain for the analysis of CTR. Along the one-dimensional spatial axis, z, the interphase density exists only for z>0. By formally interpreting z axis as the time axis, KKR is applied to obtain the relationship between $f'$ and $f''$. The structure factor of an interphase structure, $f$, satisfies the causality condition because the electron density is limited to z>0. Therefore, if the structure of the substrate is known, the direct inversion of the interphase density distribution using the KKR becomes possible.

### 2.2 Analysis of the normalized data

Normalization of the CTR data by that of a *standard state* is a way to circumvent geometric factors such as the absorption factor, polarization-Lorentz factor and the instrumental factor, etc. The standard state is chosen as close as possible to the ideally truncated substrate. However, a weak interphase structure can be present in the standard and the normalized CTR is described as $|F + f|^2/|F + f_{std}|^2$, where $f$ and $f_{std}$ are the structure factors of the unknown interphase structure and that of the standard state. Since the form, $|F + f|^2/|F|^2$, is required in Eq. (1), the obtained structure factors for inversion can be approximated by $f_{inv} = f - f_{std}$. Then, the reconstructed electron density holds the following relation:

$$\rho_{inv} = \rho - \rho_{std} \tag{4}$$

where $\rho_{inv}$, $\rho$, and $\rho_{std}$ are the inverted electron density distributions, the interphase structure of interest, and the interphase structure of the standard state, respectively. Thus, the analyzed electron density corresponds to the deviation from the standard state.

**2.3 Electron density in the negative coordinate**

The lattice expansion or contraction of the substrate can partially breaks the causality condition ($\rho = 0$ for $z < 0$) and Eq. (2) is not strictly satisfied. However, the inversion is still possible because Hilbert transformation has a property of changing the sign of the density for $z<0$ and adding to the density for $z>0$. Because of this property, the electron density can be obtained in the following relation:

$$\rho_{inv}(z) \equiv \rho_+(z) - \rho_-(-z) \tag{5}$$

where + and – denotes the densities in the positive and negative coordinates, respectively. Further detailed discussion for the sign change is given in SM.

## 3. Numerical simulation

Numerical simulations were performed to demonstrate the validity of the present method using an electron density distribution shown in Fig. 1(a). The normalized CTR intensity was calculated as defined above. The CTR structure factor, $F$, was calculated assuming point-like atoms with the lattice spacing of 2.5 Å for simplicity. Then, the normalized intensity defined in Eq. (1) was numerically solved under the constraint of Eq. (2) using the trust-region-dogleg algorithm (Powell, 1970), which typically requires a few iterations for the convergence. The reconstructed electron density is in excellent agreement with the original electron density distribution (see Fig. 1 (a)), where their discrepancy is less than 2 x $10^{-4}$ in the electron density.

The simulations were repeated with the substrate lattice planes slightly expanded. The layer expansions are defined by fractions of the lattice spacing, $\epsilon_m$ for $m$=1,2,3, as in Fig. 1 (b). The structure factor becomes:

$$f_- = \exp(\pi i l) \sum_m \{\exp[-2\pi i l(m + \epsilon_m)] - \exp(-2\pi i l m)\}, \qquad (6)$$

where $\epsilon_m$ were set to be 0.01, 0.005, and 0.001 for 1st, 2nd and 3rd layer, respectively. A lattice expansion or contraction appears as a dipole, a pair of the delta functions with opposite signs (blue dash line), in Fig. 1 (b). The noise indicated by arrows in Fig. 1 (b) are due to the limited range of the structure factor. The electron density distribution obtained by an inversion appears in the positive coordinate with its sign changed as predicted by Eq. (6). This simulation implies that the change in the substrate appears in the positive coordinate, overlapping the interphase structure. The inversion is actually a general feature of the CTR analysis. It means that one cannot judge whether a certain reconstructed electron density is from the positive or negative coordinates only from the scattering data; however, a priori knowledge about the substrate such as the existence of the lattice strain enables us to correct or interpret the results. The lattice strain can be combined to the electron density distribution in the simulation. The direct inversion of the combined density distribution is quite straightforward and given in SM.

## 4. Analysis of experimental data

The electrochemical interphase of Pt(111) single crystal immersed in a 0.1 M CsF aqueous solution was studied at the beamline 11ID-D in Advanced Photon Source in a new transmission cell geometry. The counter and reference electrodes were a Pt wire and Ag/AgCl in 3 M KCl, respectively. The transmission cell and other experimental details will be published elsewhere. The CTR profiles were measured from (0 0 0.2) to (0 0 7.5) in the hexagonal index of face-centered cubic Pt (Huang et al., 1990) using the X-ray wavelength of 0.62 Å. $(003)_h$ in the hexagonal index is equivalent to $(111)_c$ in cubic index where $a_h \equiv a_c/\sqrt{2}$, $c_h \equiv \sqrt{3}a_c$. The measured CTR data at 400 mV was chosen to be the standard CTR because it was closest to the CTR of an ideally terminated surface (You & Nagy, 1994; You et al., 1994) and close to potential of zero charge (pzc) (Petrii, 2013). Thus, the electron densities reconstructed from the normalized data correspond to the changes from those at 400 mV. The structure factor was expanded to negative $l$ region using Friedel's law before data analysis. The lattice spacing of $d_{003} = 2.2661$ Å and atomic form factor (Waasmaier et al., 1995) were used for calculating the structure factor of the bulk Pt substrate. The surface Debye-Waller factor, $\exp(-(\sigma l)^2)$, with $\sigma = 0.348$, determined by preliminary analysis, was also taken into account for the substrate. The discrete inverse Fourier transform was used to calculate the electron density after applying the linearly interpolated Hann window function (Blackman & Tukey, 1958) to the calculated structure factor.

The structure factor was retrieved from the experimental data in the same manner as in the simulations. The first two layers of the substrate are found to be strained and overlap the interphase structure at $z = d_{003}/2$ and $3d_{003}/2$ as shown in Fig. 2 (b). In addition, an indeterminate constant in $f'$ by the Hilbert transform causes a peak at the origin. These extra contributions are together described as:

$$\Delta f = f_0 + F_a(k)\exp(\pi i l/3)\sum_m\{\exp[-2\pi i(m+\epsilon_m)l/3] - \exp(-2\pi i m l/3)\}, \quad (7)$$

where $f_0$ is the indeterminate constant. Since these contributions appear as delta functions or dipoles, it is possible to distinguish them from the real interphase structure. In the present study the contributions from the strain and the constant were subtracted from the retrieved structure factor after determining $f_0$ and $\epsilon_m$ of the first two layers by minimizing $\chi \equiv \sum_j |d\rho/dx|_j$, where $j$ denotes the data index ranging from −10 to 4.5 Å (Fig. 2 (c) and Fig. S1 in SM). The structure factor of the surface strain contributes long-period oscillations. This causes the noise in the electron density obtained by the discrete Fourier transform due to the significant truncation error. Thus, it was subtracted from the structure factor in advance before the data inversion.

The electron density distribution of the interphase exhibits two sharp peaks at 2.5 and 5.2 Å and one broad peak ranging from 12 to 20 Å (Fig. 3). These peaks are mainly due to the $Cs^+$ distribution because the number of electrons of Cs (55 $e^-$/atom) is much larger than the other constituent chemical species such as F (9 $e^-$/atom) that are essentially the background density by $H_2O$ (10 $e^-$/atom). The magnitudes of the first two peaks strongly depend on the electrode potential while the broad peak does not. The first sharp peak, the second peak, and the broad peak can be interpreted as the inner Helmholtz layer, the outer Helmholtz layer, and the diffuse layer, respectively, of the Stern model (Stern, 1924). The distance of the closest peak from the electrode surface is comparable with the ionic radius of $Cs^+$. The distance of the second closest peak from the electrode surface is similar to the second water layer (Toney et al., 1994), indicating the existence of the hydrated $Cs^+$.

The positions of the sharp peaks are consistent with the molecular dynamic (MD) simulations (Spohr, 1998). The magnitude of the first peak in the MD simulation is 0.87 $e^-$ $Å^{-3}$, where the surface charge density was set to −9.9 $\mu C\ cm^{-2}$ by using 17 cations and 15 anions instead of 16 each in the cell with periodic boundary conditions. The corresponding potential was roughly estimated to −550 mV from the pzc when a constant capacitance of 18 $\mu F\ cm^{-2}$ (Anastopoulos & Papaderakis, 2014) was used for the calculation. The magnitude of the first peak in the MD simulation is comparable with the present result of 0.309 $e^-\ \text{Å}^{-3}$ at −850 mV. The discrepancy may be due to the different solution concentration; 2.2 M used in the MD calculation, while 0.1 M was used in the experiment. The small positive and negative changes in the electron density were also observed around 0.5 and 3.6 Å. These are more likely due to the uncertainties associated with limited data ranges and statistics. However, a part of them would be attributable to the distribution of water molecules because the background is assumed zero at the water density.

## 5. Conclusion

The non-iterative direct analysis method for 1D normalized CTR was proposed and its validity was demonstrated both by numerical calculations and experiments. The Hilbert transform, which is derived from the causality, provides the relationship between real and imaginary parts of a scattering factor. Thus, it is possible to determine the complex structure factor only from the scattering intensity and to reconstruct the interphase structure. Numerical simulations demonstrated that the present analysis is valid for an interphase structure in the positive coordinate defined as an outside of a substrate. The lattice strain of the substrate top layers in the negative coordinate is reconstructed but inverted to the positive coordinates with a sign change. The method was also tested experimentally for a Pt electrode immersed in 0.1 M CsF aqueous solution. The peaks in the directly reconstructed electron density are consistent with $Cs^+$ layers in the inner and outer Helmholtz planes and a long-range diffuse layer.

The present approach should be more robust and versatile in the case of weakly perturbed interfacial structures than the conventional modeling techniques used in CTR because the present method is free from the convergence problem and from inadequate initial models. It is also capable of handling the negative scattering length when the deviations from a standard density are of interest. This method is a useful addition to many powerful x-ray surface scattering tools. The concept is also applicable to x-ray reflectivity because the expression of XRR is similar to that of CTR within the kinematical approach (You, 1992). A numerical simulation for a XRR case is shown in SM.

### Acknowledgements

The work was supported by the U.S. Department of Energy (DOE), Office of Basic Energy

Science (BES), Materials Sciences and Engineering Division and use of the APS by DOE BES Scientific User Facilities Division, under Contract No. DE-AC02-06CH11357. One of the authors (TK) thanks the Japanese Society for the Promotion of Science (JSPS) for JSPS Postdoctoral Fellowships for Research Abroad. The work at RIT was supported by the Research Corporation for Science Advancement (RCSA) through a Cottrell College Science.

**Figures**

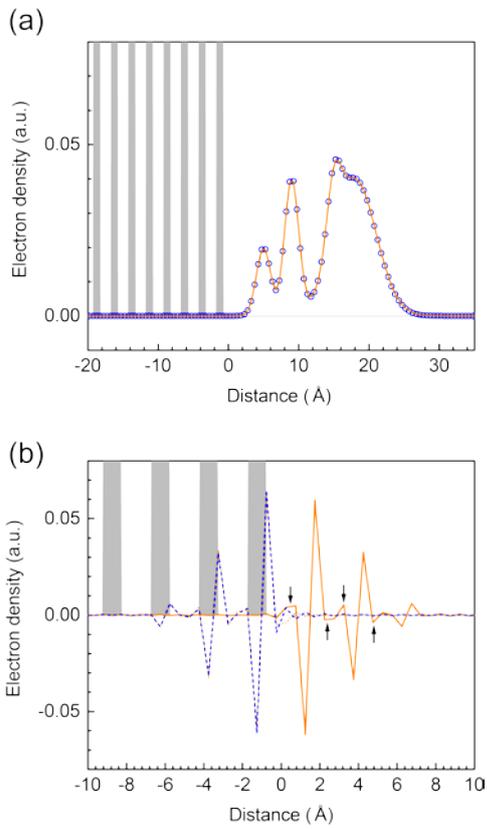

Fig. 1 (a) A simulated electron density distribution (blue circle) and the reconstructed distribution by analyzing the scattering data (orange solid line). The substrate crystal lattice is depicted as gray vertical bars. (b) Simulated lattice expansion (blue dashed line), the reconstructed profile by analyzing the scattering profile (orange solid line) and that of the flipped profile (orange dotted line). Arrows indicate the noise derived from the Fourier transform of the limited range of the structure factor.

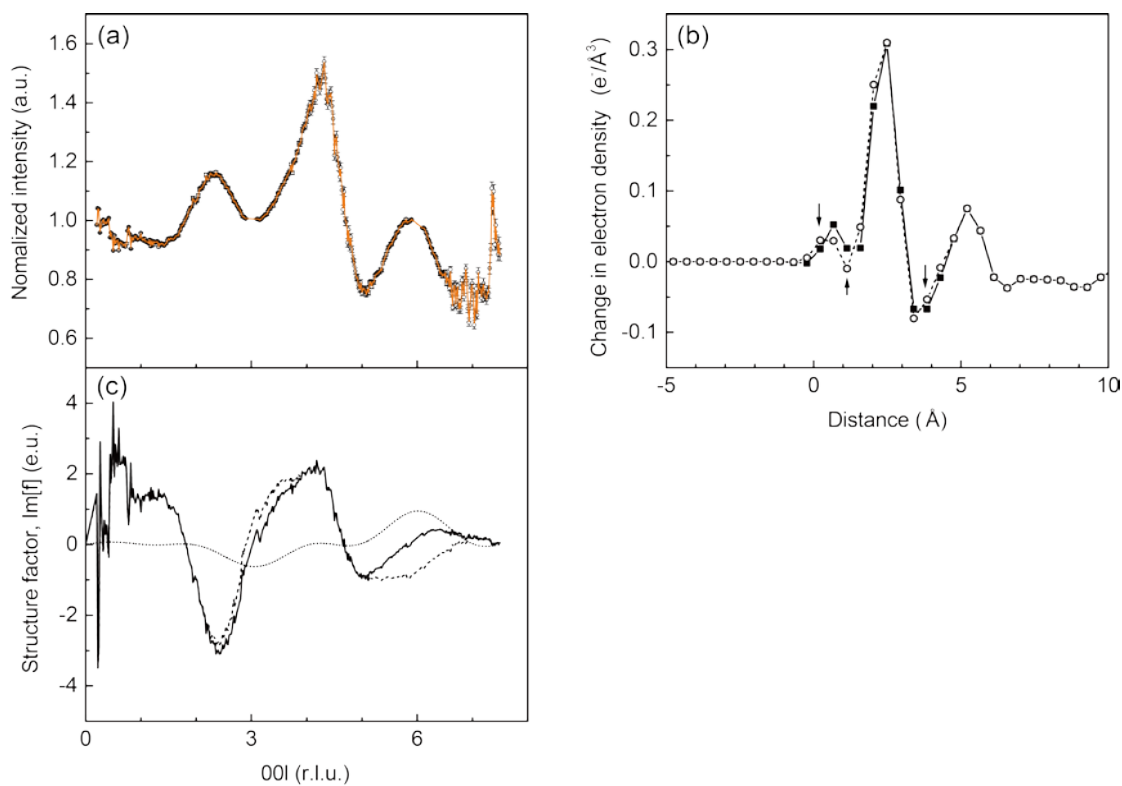

Fig. 2 (a) The normalized CTR profile obtained at −850 mV vs Ag/AgCl (black circles with error bars) and that obtained from the analyzed structure factor (orange solid line). (b) The electron density distribution near the electrode surface with (solid line with squares) and without (dashed line with circles) the subtraction of the lattice expansion of the substrate. The Pt(111) top layer is at −1.133 Å (not shown). Arrows indicate the electron density changes derived from the substrate and analytical error. (c) Imaginary parts of the structure factor: The solid, dotted, and dashed lines represent the contributions directly obtained, from the experimental data, only from the lattice strain, and only from the interphase structure, respectively.

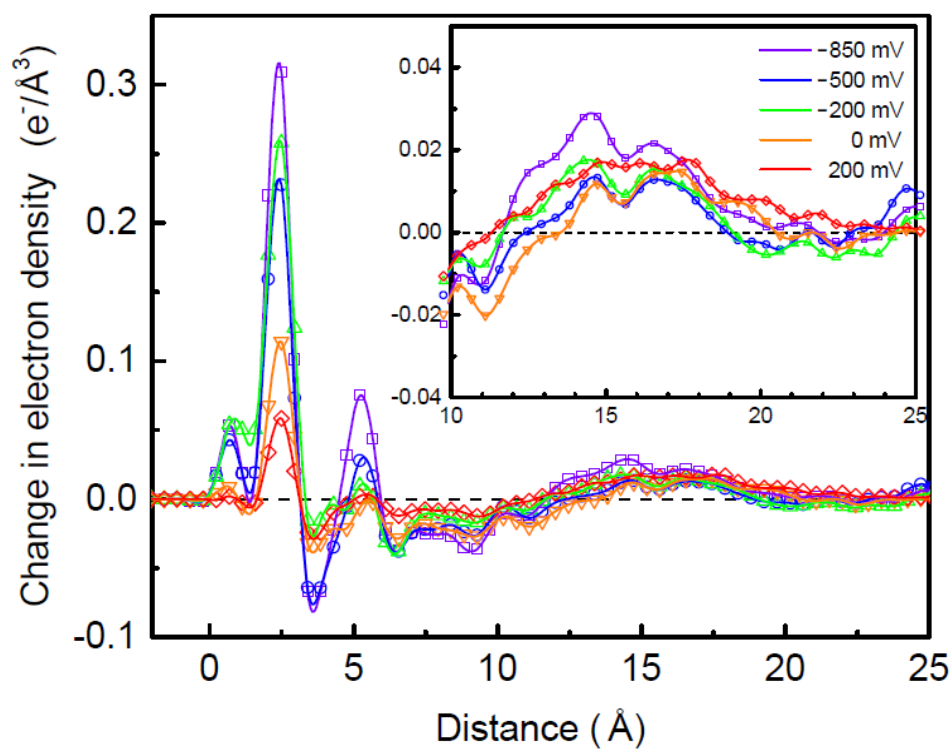

Fig. 3 The electron density profiles directly inverted at various electrode potential in 0.1 M CsF electrolyte. Inset: the range from 10 to 25 Å is magnified. The statistical errors are approximately the size of symbols. The cubic-spline curves are guides to the eye. The Pt(111) top layer is at −1.133 Å.